\documentclass[twocolumn,preprintnumbers,superscriptaddress,nofootinbib,aps,prd,floatfix]{revtex4}

\pdfoutput=1
\usepackage{graphicx}
\usepackage{latexsym}
\usepackage{amsfonts}
\usepackage{amssymb}
\usepackage{amsmath}
\usepackage{slashed}
\usepackage{feynmp}
\usepackage{hyperref}
\usepackage{url}
\usepackage{color}
\usepackage{cancel}
\usepackage{multirow,array}
 \usepackage{float}
 \usepackage{subfigure}

\begin{document}
\def\contentsname{{\normalsize Content}}
\def\tablename{Table}
\def\figurename{Figure}

\def\pveto{P_\text{veto}}
\def\nj{n_\text{jets}}
\def\meff{m_\text{eff}}
\def\ptmin{p_T^\text{min}}
\def\gtot{\Gamma_\text{tot}}
\def\as{\alpha_s}
\def\az{\alpha_0}
\def\gz{g_0}
\def\w{\vec{w}}
\def\sdag{\Sigma^{\dag}}
\def\s{\Sigma}
\newcommand{\psib}{\overline{\psi}}
\newcommand{\Psib}{\overline{\Psi}}
\newcommand\one{\leavevmode\hbox{\small1\normalsize\kern-.33em1}}
\newcommand{\Mpl}{M_\mathrm{Pl}}
\newcommand{\p}{\partial}
\newcommand{\mat}{\mathcal{M}}
\newcommand{\lag}{\mathcal{L}}
\newcommand{\ord}{\mathcal{O}}
\newcommand{\ope}{\mathcal{O}}
\newcommand{\qqquad}{\qquad \qquad}
\newcommand{\qqqquad}{\qquad \qquad \qquad}

\newcommand{\qb}{\bar{q}}
\newcommand{\matx}{|\mathcal{M}|^2}
\newcommand{\really}{\stackrel{!}{=}}
\newcommand{\msbar}{\overline{\text{MS}}}
\newcommand{\qns}{f_q^\text{NS}}
\newcommand{\lqcd}{\Lambda_\text{QCD}}
\newcommand{\met}{\slashchar{E}_T}
\newcommand{\pmiss}{\slashchar{\vec{p}}_T}

\newcommand{\sq}{\tilde{q}}
\newcommand{\go}{\tilde{g}}
\newcommand{\st}[1]{\tilde{t}_{#1}}
\newcommand{\stb}[1]{\tilde{t}_{#1}^*}
\newcommand{\nz}[1]{\tilde{\chi}_{#1}^0}
\newcommand{\cp}[1]{\tilde{\chi}_{#1}^+}
\newcommand{\cm}[1]{\tilde{\chi}_{#1}^-}
\newcommand{\CP}{CP}

\providecommand{\mg}{m_{\tilde{g}}}
\providecommand{\mst}[1]{m_{\tilde{t}_{#1}}}
\newcommand{\msn}[1]{m_{\tilde{\nu}_{#1}}}
\newcommand{\mch}[1]{m_{\tilde{\chi}^+_{#1}}}
\newcommand{\mne}[1]{m_{\tilde{\chi}^0_{#1}}}
\newcommand{\msb}[1]{m_{\tilde{b}_{#1}}}
\newcommand{\vsm}{\ensuremath{v_{\rm SM}}}

\newcommand{\mev}{{\ensuremath\rm MeV}}
\newcommand{\gev}{{\ensuremath\rm GeV}}
\newcommand{\tev}{{\ensuremath\rm TeV}}
\newcommand{\fb}{{\ensuremath\rm fb}}
\newcommand{\ab}{{\ensuremath\rm ab}}
\newcommand{\pb}{{\ensuremath\rm pb}}
\newcommand{\sign}{{\ensuremath\rm sign}}
\newcommand{\iab}{{\ensuremath\rm ab^{-1}}}
\newcommand{\ifb}{{\ensuremath\rm fb^{-1}}}
\newcommand{\ipb}{{\ensuremath\rm pb^{-1}}}

\def\slashchar#1{\setbox0=\hbox{$#1$}           
   \dimen0=\wd0                                 
   \setbox1=\hbox{/} \dimen1=\wd1               
   \ifdim\dimen0>\dimen1                        
      \rlap{\hbox to \dimen0{\hfil/\hfil}}      
      #1                                        
   \else                                        
      \rlap{\hbox to \dimen1{\hfil$#1$\hfil}}   
      /                                         
   \fi}
\newcommand{\dslash}{\slashchar{\partial}}
\newcommand{\Dslash}{\slashchar{D}}

\newcommand{\eg}{\textsl{e.g.}\;}
\newcommand{\ie}{\textsl{i.e.}\;}
\newcommand{\etal}{\textsl{et al}\;}

\setlength{\floatsep}{0pt}
\setcounter{topnumber}{1}
\setcounter{bottomnumber}{1}
\setcounter{totalnumber}{1}
\renewcommand{\topfraction}{1.0}
\renewcommand{\bottomfraction}{1.0}
\renewcommand{\textfraction}{0.0}
\renewcommand{\thefootnote}{\fnsymbol{footnote}}

\newcommand{\rig}{\rightarrow}
\newcommand{\lrig}{\longrightarrow}
\renewcommand{\d}{{\mathrm{d}}}
\newcommand{\be}{\begin{eqnarray*}}
\newcommand{\ee}{\end{eqnarray*}}
\newcommand{\gl}[1]{(\ref{#1})}
\newcommand{\ta}[2]{ \frac{ {\mathrm{d}} #1 } {{\mathrm{d}} #2}}
\newcommand{\bee}{\begin{eqnarray}}
\newcommand{\eee}{\end{eqnarray}}
\newcommand{\beeq}{\begin{equation}}
\newcommand{\eeeq}{\end{equation}}
\newcommand{\mc}{\mathcal}
\newcommand{\mr}{\mathrm}
\newcommand{\ep}{\varepsilon}
\newcommand{\emt}{$\times 10^{-3}$}
\newcommand{\emfo}{$\times 10^{-4}$}
\newcommand{\emfi}{$\times 10^{-5}$}

\newcommand{\revision}[1]{{\bf{}#1}}

\newcommand{\hzero}{h^0}
\newcommand{\Hzero}{H^0}
\newcommand{\Azero}{A^0}
\newcommand{\PHiggs}{H}
\newcommand{\PW}{W}
\newcommand{\PZ}{Z}

\newcommand{\sw}{\ensuremath{s_w}}
\newcommand{\cw}{\ensuremath{c_w}}
\newcommand{\swd}{\ensuremath{s^2_w}}
\newcommand{\cwd}{\ensuremath{c^2_w}}

\newcommand{\mhhd}{\ensuremath{m^2_{\Hzero}}}
\newcommand{\mhh}{\ensuremath{m_{\Hzero}}}
\newcommand{\mlhd}{\ensuremath{m^2_{\hzero}}}
\newcommand{\Mlh}{\ensuremath{m_{\hzero}}}
\newcommand{\mad}{\ensuremath{m^2_{\Azero}}}
\newcommand{\mhpd}{\ensuremath{m^2_{\PHiggs^{\pm}}}}
\newcommand{\mhp}{\ensuremath{m_{\PHiggs^{\pm}}}}

 \newcommand{\sa}{\ensuremath{\sin\alpha}}
 \newcommand{\ca}{\ensuremath{\cos\alpha}}
 \newcommand{\cad}{\ensuremath{\cos^2\alpha}}
 \newcommand{\sad}{\ensuremath{\sin^2\alpha}}
 \newcommand{\sbd}{\ensuremath{\sin^2\beta}}
 \newcommand{\cbd}{\ensuremath{\cos^2\beta}}
 \newcommand{\cb}{\ensuremath{\cos\beta}}
 \renewcommand{\sb}{\ensuremath{\sin\beta}}
 \newcommand{\tanbd}{\ensuremath{\tan^2\beta}}
 \newcommand{\cotbd}{\ensuremath{\cot^2\beta}}
 \newcommand{\tanb}{\ensuremath{\tan\beta}}
 \newcommand{\tb}{\ensuremath{\tan\beta}}
 \newcommand{\cotb}{\ensuremath{\cot\beta}}

\newcommand{\GeV}{\ensuremath{\rm GeV}}
\newcommand{\MeV}{\ensuremath{\rm MeV}}
\newcommand{\TeV}{\ensuremath{\rm TeV}}

\title{Constraining the Strength and CP Structure of Dark Production at the LHC: \\
the Associated Top-Pair Channel}

\author{Matthew R.~Buckley}
\affiliation{Department of Physics and Astronomy, Rutgers University, Piscataway, NJ 08854, USA}
\author{Dorival Gon\c{c}alves}
\affiliation{Institute for Particle Physics Phenomenology, Department of Physics, Durham University, UK}

\begin{abstract}
We consider the production of dark matter in association with a pair of top quarks, mediated by a scalar 
or pseudoscalar particle in a generic Simplified Model. We demonstrate that the difference of azimuthal 
angle between the two leptons $\Delta \phi_{\ell\ell}$, in the dileptonic top decay mode, can directly probe the 
CP-properties of the mediator. We estimate the constraints to strength and CP-structure of dark matter production
for  these well-motivated Simplified Models  from the LHC Run II. 
\end{abstract}

\maketitle

\section{Introduction \label{sec:intro}}

The existence of dark matter is solid evidence for physics beyond the Standard Model. Though currently known only through gravitational interactions, there are well-motivated theoretical reasons to expect dark matter to have significant interactions with the known particles. If dark matter is a particle that was in thermal equilibrium in the early Universe, it must have a sufficiently large annihilation cross section with some other set of particles, in order reduce its thermal relic abundance at least to the level of the observed dark matter density. If the only form of dark matter are these thermal relics, then the required thermally averaged cross section, $\langle \sigma v\rangle \sim 3 \times 10^{-26}~$cm$^3$/s, is strongly suggestive of the scale of weak-force particles. However, the observation that some interaction is necessary to reduce the number density is much more general and can apply to non-thermal dark matter, such as that found in Asymmetric Dark Matter models \cite{Buckley:2011kk}.

Given this motivation, there is a great deal of interest, both theoretical and experimental, in the search for dark matter particles in particle physics experiments. In particular, given an approximately weak-scale-strength interaction so as not to over-close the Universe with dark matter, the multipurpose experiments ATLAS and CMS at the Large Hadron Collider (LHC), have the potential to pair-produce dark matter and detect it via events with an excess of missing transverse energy ($\slashed{E}_T$) \cite{Birkedal:2004xn,Feng:2005gj,Beltran:2008xg,Konar:2009ae,Beltran:2010ww,Bai:2010hh,Rajaraman:2011wf,Fox:2011pm,Bai:2012xg,Fox:2012ee,Carpenter:2012rg}. 

Recently, it has been recognized that accurate simulation of dark matter production at the LHC requires the inclusion of the particle(s) that mediate the interaction with the Standard Model for a significant range of the parameter space~\cite{Fox:2011fx,Shoemaker:2011vi,Weiner:2012cb,Busoni:2013lha,Buchmueller:2013dya,Buchmueller:2014yoa,Busoni:2014sya,Busoni:2014haa}. In order to avoid a loss of generality by focusing on a single UV-complete model such as supersymmetry, a framework has been constructed of {\it Simplified Models} \cite{Alwall:2008ag,Alves:2011wf,Goodman:2011jq,Jacques:2015zha,Godbole:2015gma,Abdallah:2015ter,Abercrombie:2015wmb}, in which the only new particles considered are the dark matter itself and the mediator that connects the dark sector with the visible. Of these {\it Simplified Models}, those with dark matter interactions mediated by a new scalar or pseudoscalar spin-0 particle interacting with the Standard Model quarks and leptons \cite{Buckley:2014fba,Haisch:2012kf,Haisch:2013ata,Haisch:2013fla,Crivellin:2014qxa,Ghorbani:2014qpa,Backovic:2015soa,Harris:2015kda,Harris:2014hga,Mattelaer:2015haa,Fan:2015sza,Khoze:2015sra,Haisch:2015ioa,Berlin:2015wwa} have particular interest, and are the focus of this paper.

In order to avoid flavor constraints, this new mediator $\Phi$ (either a scalar $H$ or pseudoscalar $A$) is assumed to couple to the Standard Model quarks via Minimal Flavor Violation (MFV) \cite{D'Ambrosio:2002ex}, in which the coupling is proportional to the Higgs Yukawas. Such interactions are suggestive of the mediator being the heavy Higgs or pseudoscalar Higgs of a non-minimal Higgs sector, or even the 125~GeV Higgs itself ({\it i.e.}~Higgs Portal dark matter \cite{Patt:2006fw,Djouadi:2011aa,Djouadi:2012zc,Goncalves:2015mfa,Corbett:2015ksa,Bernaciak:2014pna,Englert:2013gz,Endo:2014cca,Englert:2011yb,Craig:2014lda,Craig:2015jba}). Under these assumptions, $\Phi$ couples predominantly to the heaviest Standard Model fermion, the top quark. Dark Matter would therefore be predominantly pair-produced either through a top quark loop-induced coupling to gluons, or in association with a top quark pair. Both mechanisms have been studied in the context of {\it Simplified Models}, and in general they yield competitive bounds on the strength of the mediator couplings  to the quarks and to dark matter~\cite{Aad:2014vea,Buckley:2014fba,Haisch:2015ioa}.

In this paper, we consider how to distinguish the structure of dark matter production at the LHC.  We exploit the sensitivity of the azimuthal angular  correlation
between the two leptons $\Delta \phi_{\ell\ell}$  arising from the tops in the associated top pair channel. 
This experimentally clean observable was previously employed by the authors in the context of direct CP-measurements of the 125~GeV Higgs boson's 
coupling to top quarks~\cite{Buckley:2015vsa}, and was motivated by similar techniques developed for studying the Higgs 
sector~\cite{Ellis:2013yxa,Boudjema:2015nda,tth_NLO,Biswas:2014hwa,khabibi,Kolodziej:2015qsa}. 
This variable contains information on the CP-structure of the mediator couplings to quarks, and so -- in the case of a discovery -- 
can be used to distinguish between the CP-even scalar and CP-odd pseudoscalar mediators. The sensitivity to the CP-structure of the mediator coupling
is enhanced when the mediator is produced at large transverse momentum. This is the kinematic regime required by the background-reducing selection 
on $\slashed{E}_T$, providing a useful synergy between the requirements for discovery and the ability to measure the properties of the new physics.

This paper is organised as follows. In Section~\ref{sec:theory}, we review the spin-0 {\it Simplified Model} formalism and the analytical structure
underlying the $\Delta\phi_{\ell\ell}$ CP sensitivity.  In Section~\ref{sec:simulation}, we describe our LHC simulation techniques. The limits on strength
and CP-structure of Simplified Models are presented in~Section~\ref{sec:results}. 

\section{Simplified Models and CP-Measurement \label{sec:theory}}
\subsection{Dark Matter Simplified Models}

We adopt the {\it Simplified Model} framework for spin-0 mediators, assuming fermionic dark matter $\chi$, as described in 
Refs.~\cite{Buckley:2014fba,Abdallah:2015ter,Abercrombie:2015wmb}. This is a five-parameter model. There are two 
masses: that of the dark matter $m_\chi$, and the mediator $m_{H(A)}$, and two couplings: the dark matter-mediator 
coupling $g_\chi$, the universal mediator-Standard Model fermion coupling $g_v$. Finally, there is the mediator width
$\Gamma_{H(A)}$, which may be considered a free parameter to allow for additional new physics beyond the purview 
of the {\it Simplified Model}. For simplicity, in this work we will consider bounds on the visible coupling $g_v$ for on-shell
production of dark matter ($m_\chi < m_{H(A)}/2$), assuming a 100\% branching ratio of $\Phi$ to dark matter. 
In this case, the detector bounds on the model are independent of $m_\chi$ for $m_{H(A)}$ within the kinematic
reach of the LHC. Under the assumption of MFV, the Lagrangian for the interaction terms between a spin-0 mediator 
$\Phi$, the dark matter $\chi$, and the Standard Model fermions $f$ is
\begin{eqnarray}
{\cal L} & = & - g_v \frac{m_f}{v} \bar{f} (\cos \alpha + i \gamma_5 \sin\alpha)f \Phi \\ 
& & - g_\chi \bar{\chi} (\cos \alpha + i \gamma_5 \sin\alpha) \chi \Phi. \nonumber
\end{eqnarray}
Here, $\alpha$ is a CP-phase, with $\alpha = 0$ corresponding to the CP-even scalar mediator $H$, and $\alpha = \pi/2$ corresponding to the pseudoscalar mediator $A$.

This Lagrangian will generate a $\Phi$-gluon-gluon coupling at one-loop, due primarily to the $\Phi$ coupling to top quarks. 
This allows for ``monojet'' searches \cite{Goodman:2010yf,Goodman:2010ku,Beltran:2010ww,Rajaraman:2011wf,Fox:2011pm,Fox:2012ee,ATLASmonojet,Khachatryan:2014rra}
in the $gg \to (\Phi \to \chi\bar{\chi})+$jets channel. However, a channel with similar sensitivity is dark matter produced through the 
$\Phi$ particle in association with a pair of top quarks \cite{Aad:2014vea}. We will consider the dileptonic top associated channel
in this paper. The details of the event selection will be considered in Section~\ref{sec:simulation}; at this point the key factor is that 
separating the associated production of tops and dark matter from the dileptonic top background requires events with large missing
transverse energy, which translates into requiring the $\Phi$ to be produced with large $p_T$. 
\begin{figure*}[th!]
\centering
\includegraphics[width=0.8\columnwidth]{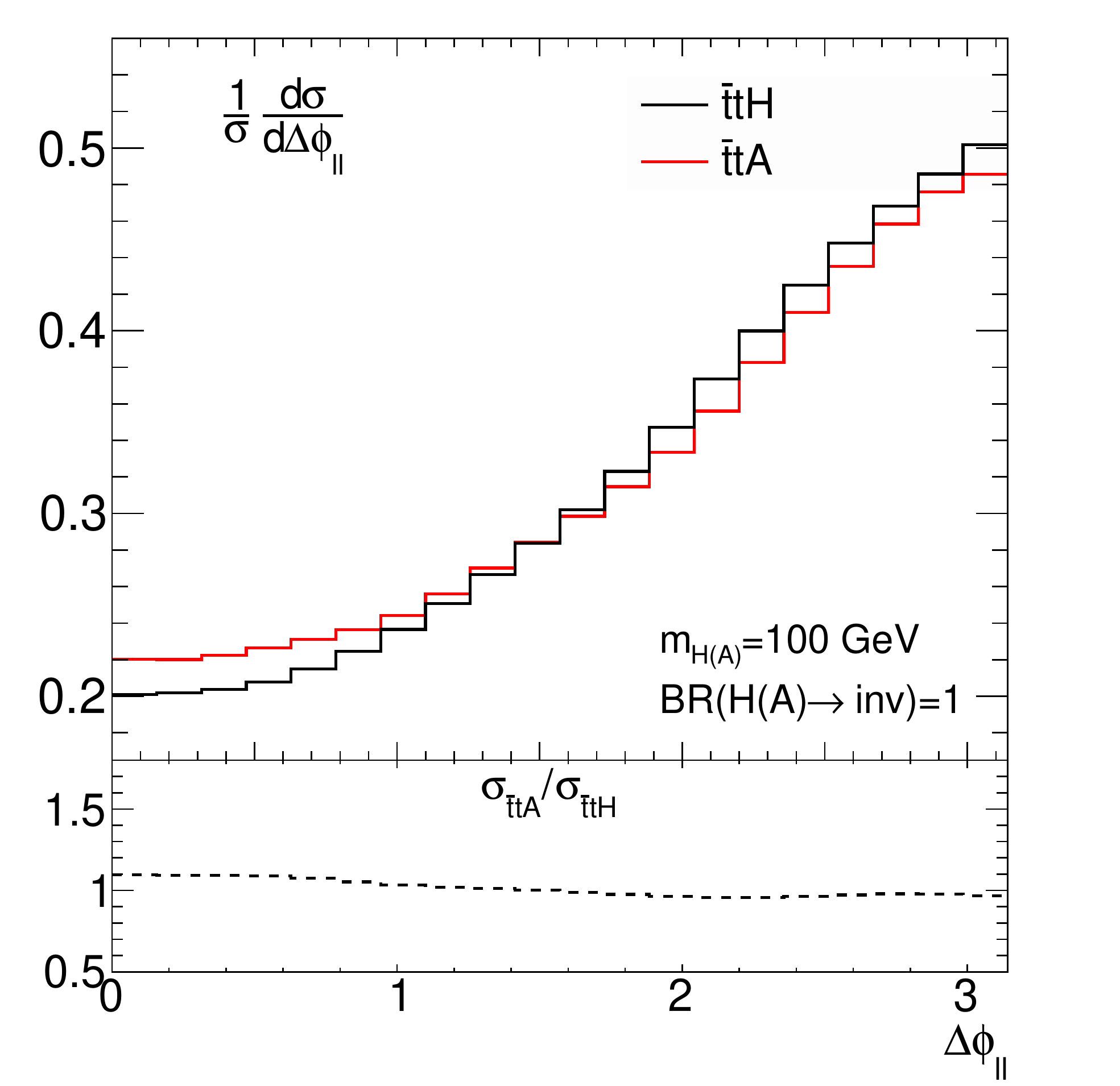}
\hspace{1cm}
\includegraphics[width=0.8\columnwidth]{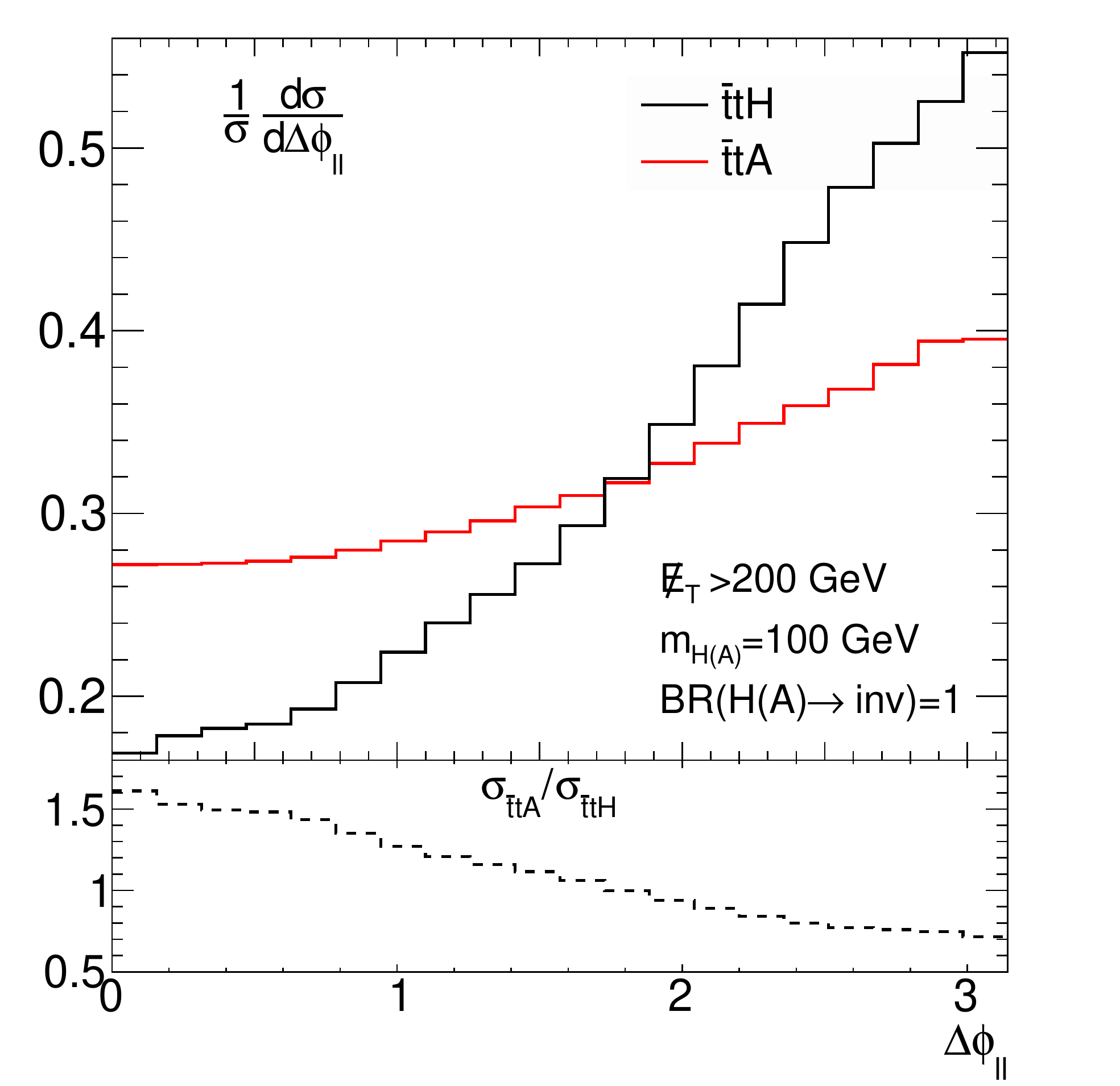}
\caption{Normalized differential distribution of $\Delta\phi_{\ell\ell}$ for signal $t\bar{t}H(A)$ events at the 13~TeV LHC, for the
CP-even $H$ and CP-odd $A$ mediators. We assume $m_{H(A)} = 100$~GeV and $m_\chi <m_{H(A)}/2 $ with $BR(H(A)\rightarrow \rm{inv.})=1$. 
Left: Distributions without missing transverse energy $\slashed{E}_T$ selection. Right: Distributions with $\slashed{E}_T > 200$~GeV selection. 
\label{fig:deltaphill}}
\end{figure*}

\subsection{CP Properties of $\Delta \phi_{\ell\ell}$}

The experimentally accessible variable of interest is $\Delta\phi_{\ell\ell}$, the azimuthal angle in the lab-frame between the leptons  from dileptonic tops in $pp \to t\bar{t}\Phi \to (\ell^+\nu b)(\ell^-\bar{\nu}\bar{b})(\chi\bar{\chi})$ events. This experimentally clean observable works as proxy for the azimuthal angle $\Delta\phi_{tt}$, the azimuthal angle between the tops in the event, which cannot be reconstructed without hadronic decays of the tops resulting in large experimental uncertainties. The correlation between $\Delta\phi_{tt}$ and $\Delta\phi_{\ell\ell}$ was discussed by the authors in Ref.~\cite{Buckley:2015vsa}. Here, we briefly recapitulate the arguments, demonstrating that $\Delta\phi_{tt}$ contains information about the CP phase-angle $\alpha$. As top decay emits leptons preferentially along the top momentum axis \cite{Mahlon:1995zn,Peskin}, the easily constructed variable $\Delta\phi_{\ell\ell}$ allows this information to be accessed experimentally.\medskip

As in Ref.~\cite{Buckley:2015vsa}, we specialize to particular subset of $t\bar{t}\Phi$ production: quark-initiated production of mixed-helicity tops, {\it i.e.}~$q\bar{q} \to t_{L/R}\bar{t}_{R/L}\Phi$. Here, we use helicity definitions as in Ref.~\cite{helas}. These final states transform into themselves under the application of CP symmetry, and the quark initial states become more important at high transverse momenta at the LHC due to falling gluon Parton Distribution Function (PDF). The matrix element for this restricted subset of final states can be written as
\begin{eqnarray*}
{\cal M}  & \propto & \frac{m_t\left[\bar{v}_2 \gamma^\mu u_1\right]\left[\bar{u}_{t}P_{L/R} {\cal A} \gamma_\mu P_{R/L}v_{\bar{t}} \right]}{[q_1+q_2]^2\left[m_{H(A)}^2+2k_t\cdot p\right] \left[m_{H(A)}^2+2k_{\bar{t}}\cdot p\right]}, \\
{\cal A} & = & \left[\tfrac{m_{H(A)}^2}{2}+(k_t+k_{\bar{t}})\cdot p \right]c_\alpha-i\left[ (k_t-k_{\bar{t}})\cdot p\right]\gamma_5s_\alpha,
\end{eqnarray*}
where $c_\alpha =\cos\alpha$, $s_\alpha = \sin\alpha$, $q_1$ and $q_2$ are the momenta of the incoming quarks (with spinors $u_1$ and $v_2$), the outgoing tops have momenta $k_t$ and $k_{\bar{t}}$, and the outgoing $\Phi$ has momentum $p$. 

When the $\Phi$ particle has significant transverse momentum, the coefficient of the CP-even $\cos\alpha$ piece of ${\cal A}$ is proportional to cosines of $\Delta\phi_{tt}$, while the CP-odd $\sin\alpha$ term is proportional to sines of $\Delta\phi_{tt}$ \cite{Buckley:2015vsa}. Thus, for these mixed helicity states, at large $p_T$, events containing a CP-even $H$ particle have an excess of events near $\Delta\phi_{tt} = 0$ and an deficit near $\Delta\phi_{tt} = \pi$. This pattern is reversed for a CP-odd $A$ mediator. As demonstrated in Ref.~\cite{Buckley:2015vsa}, this behavior survives in the experimentally accessible variable $\Delta\phi_{\ell\ell}$, where the azimuthal angle of the leptons is used as a proxy for the top/antitop angle. In Figure~\ref{fig:deltaphill}, we show the differential distributions of $\Delta\phi_{\ell\ell}$ for the CP-even $H$ and CP-odd $A$, assuming $m_{H(A)}=100$~GeV. In the left-hand panel, we do not apply a cut on $\slashed{E}_T$, while on the right we require $\slashed{E}_T > 200$~GeV. This not only forces the surviving events into the configuration where the CP-structure of the term ${\cal A}$ is made manifest, it also boosts the relative admixture of $q\bar{q}$ initial states, and the contributions from mixed-helicity top pairs. This results in the significant excess of events near $\Delta\phi_{\ell\ell} \sim 0$ in the CP-odd case as compared to the CP-even.

\section{Simulation and Event Selection \label{sec:simulation}}

\begin{figure}[!hb]
\centering
\includegraphics[width=0.95\columnwidth]{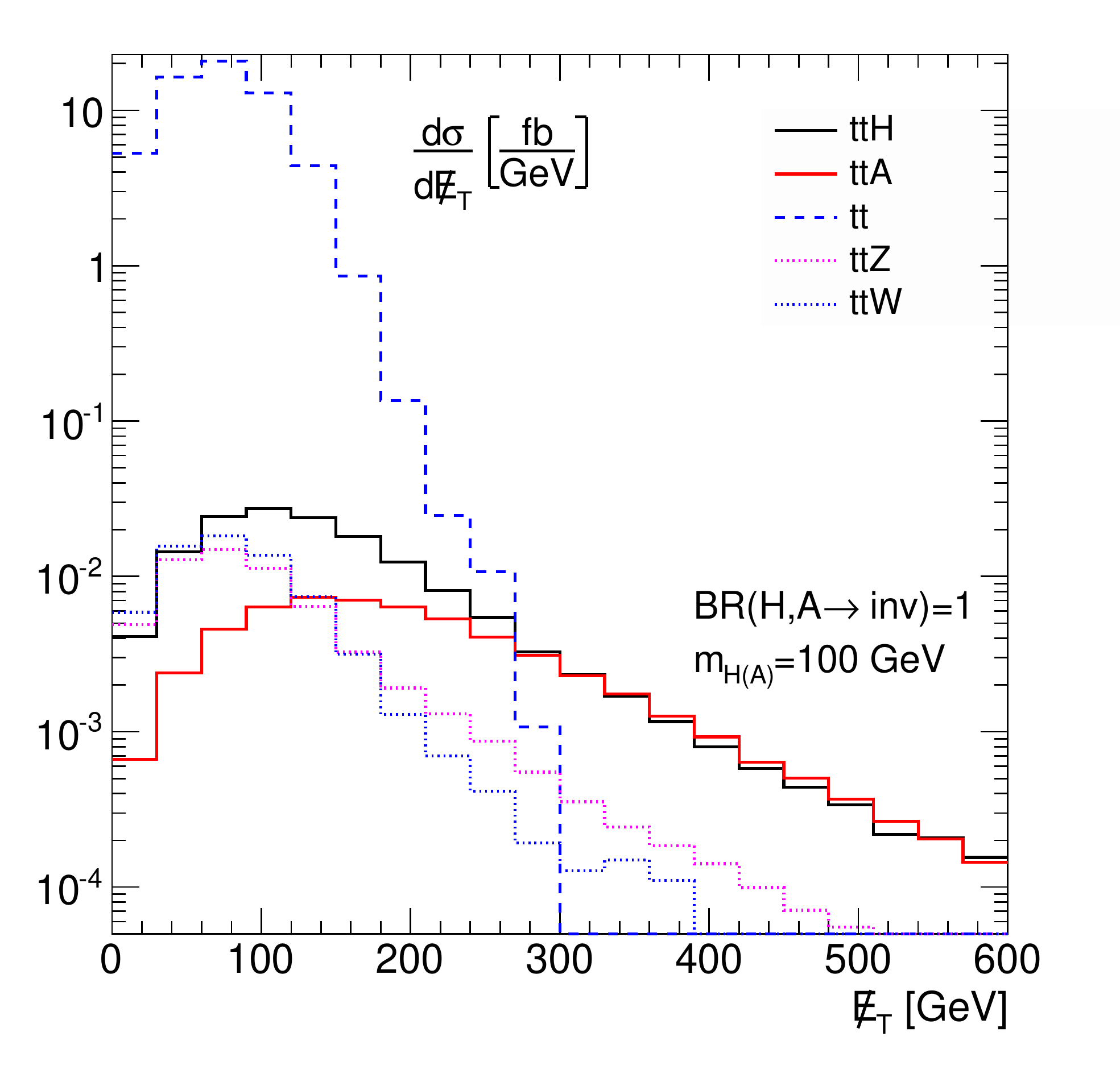}
\caption{Missing energy distribution $\slashed{E}_T$ for the signal processes $t\bar{t}H$ and $t\bar{t}A$ 
as well as the primary backgrounds. 
Signal events assume  $m_{H(A)} = 100$~GeV, $g_v = 1$ and
$m_\chi < m_{H(A)}/2 $ with $BR(H,A\rightarrow \rm{inv.})=1$.
\label{fig:etmiss}}
\end{figure}

Having developed a variable that is sensitive to the CP structure from the dark matter mediator, we first present an 
analysis strategy to measure the signal strengths at the Run II LHC, and then  directly distinguish the CP structure. 

The signal samples are $pp\rightarrow t \bar{t} \Phi$ ($\Phi = H$ or $A$) with fully dileptonic top decays and mediator (pseudo-)scalar decays 
to invisibles. We require two opposite-sign leptons and two bottom tagged jets. The major backgrounds for this 
signature are top pair $t\bar{t}$ production and associated top pair with Z or W gauge bosons, $t\bar{t}Z$ and $t\bar{t}W$. 

We simulate our signal samples  with \textsc{MadGraph5+Pythia8}~\cite{mg5,pythia8} and the backgrounds with
\textsc{Sherpa+OpenLoops}~\cite{sherpa,sherpa2,sherpa3,sherpa4,sherpa5,openloops}. All samples are generated including higher order QCD 
effects with the \textsc{MC@NLO} algorithm~\cite{mcatnlo} accounting for hadronization and underlying event effects.  
We use  \textsc{MadSpin}~\cite{madspin} and  the respective  \textsc{Sherpa}~\cite{sherpa_spin} module
to restore spin correlations between production and decays for the tops and gauge bosons.\medskip

In our analysis we require two isolated  leptons with $p_{Tl}>20~\gev$  and $|\eta_\ell|<2.5$.  Leptons are isolated
if  less than $20\%$ of the transverse energy deposited in the lepton radius $R=0.2$ corresponds to hadronic activity. 
The hadronic part of the event is reconstructed using \textsc{Fastjet}~\cite{fastjet,fastjet2} via the anti-$k_T$ jet  algorithm with radius $R=0.4$. 
Jets are defined with $p_{Tj}>30~\gev$ and $|\eta_j|<2.5$.   We demand two $b$-tagged jets to suppress possible extra sources 
of backgrounds. We assume a $b$-tagging efficiency of $70\%$ and a mistag rate of $1\%$. 

The main variable used to distinguish signal and background in existing searches is the missing transverse energy $\slashed{E}_T$. 
In Figure~\ref{fig:etmiss}, we show the $\slashed{E}_T$ distribution of both the scalar and the pseudoscalar mediator 
production with  dileptonic tops, assuming $m_{H(A)} = 100$~GeV, $g_v = 1$ and $m_\chi <m_{H(A)}/2$. It is clear that 
a large $\slashed{E}_T$ selection can be used to enhance the sensitivity to the signal sample and deplete the $t\bar{t}$
background. Moreover, we notice that pseudoscalar mediators lose a smaller fraction of the rate with large $\slashed{E}_T$ 
selections than the scalar case. We observe that the azimuthal correlation between the most energetic jet and lepton helps in the 
$t\bar{t}$ background suppression, so we demand $\Delta\phi_{j\ell}<2$.

\section{Results and Conclusion \label{sec:results}}
\begin{figure*}[!tb]
\begin{center}
\includegraphics[width=0.9\columnwidth]{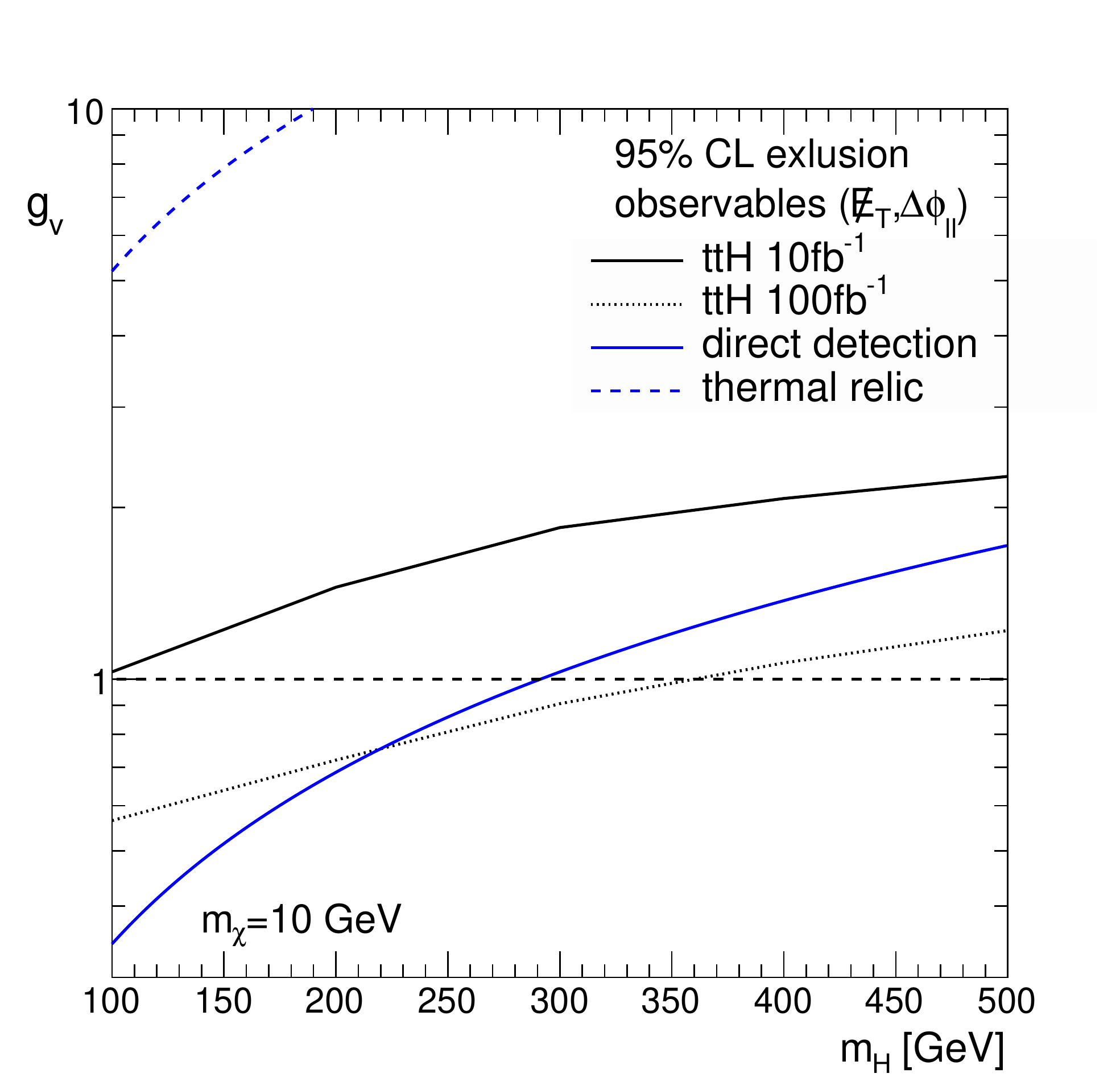}
\includegraphics[width=0.9\columnwidth]{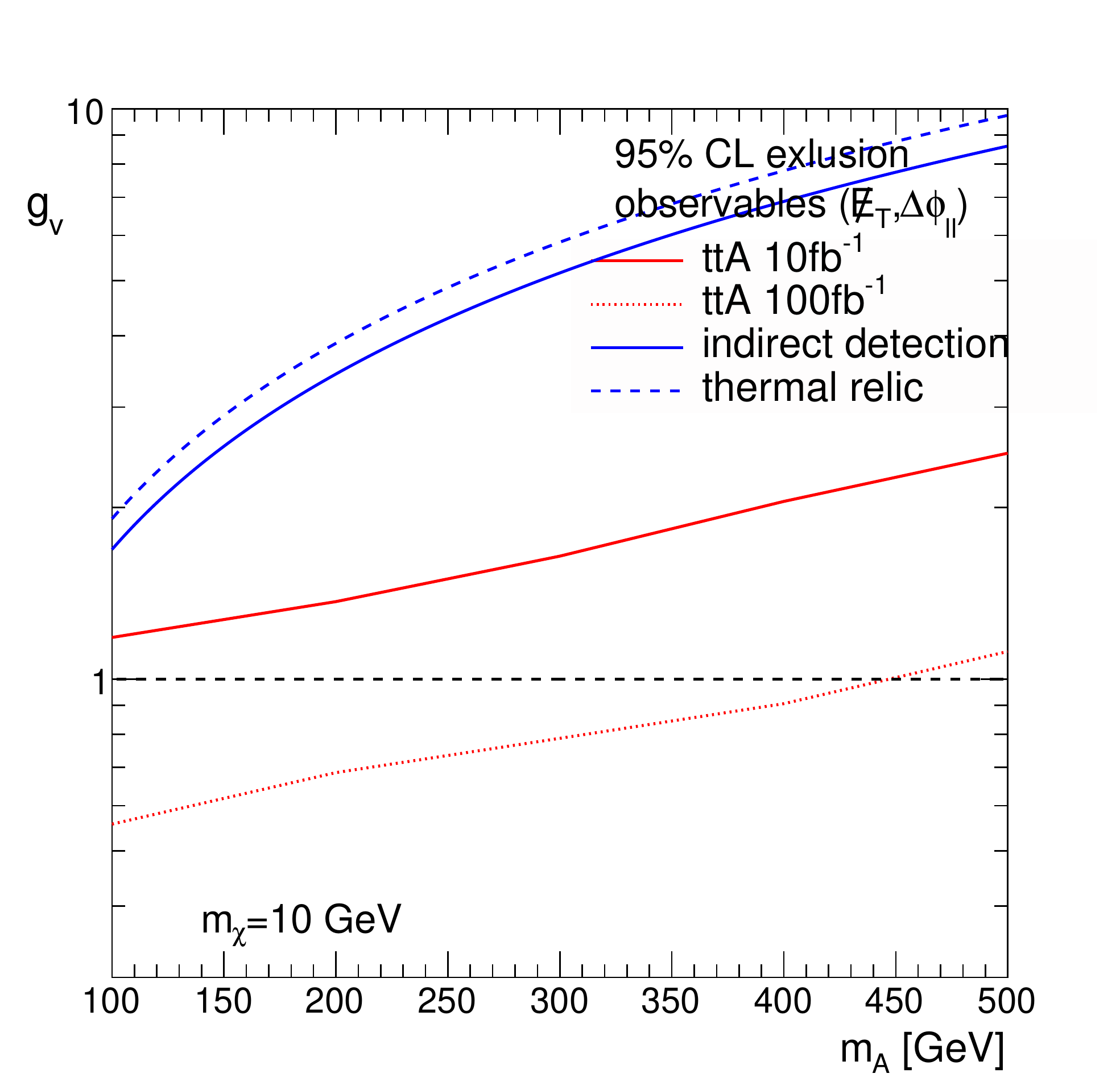}
\end{center}
\caption{95\% CL upper limit on the mediator coupling to tops $g_v$ as a function of the mediator mass $m_{H(A)}$ for scalars (left) and pseudoscalars (right).
The log-likelihood test is based on the two dimensional distributions $(\slashed{E}_T,\Delta\phi_{\ell\ell})$
for 10~fb$^{-1}$ and 100~fb$^{-1}$ of LHC data with $BR(H,A\rightarrow \rm{inv.})=1$. The scalar mediators also have 95\% CL upper limits on $g_v$ from LUX direct detection bounds \cite{Akerib:2013tjd}, while the pseudoscalar mediators are constrained by the {\it Fermi}-LAT stacked dwarf galaxy search \cite{Ackermann:2015zua}. In both cases, the nominal value of $g_v$ required for a thermal relic is also shown. 
\label{fig:constrain}}
\end{figure*}

We perform a 2-dimensional binned log-likelihood analysis, using both $\slashed{E}_T$ and $\Delta\phi_{\ell\ell}$ to distinguish signal from background. In Figure~\ref{fig:constrain}, we show the expected 95\% CL limit at the 13~TeV LHC on the mediator coupling to tops $g_v$ as a function of the mediator mass, for both the scalar and pseudoscalar. We assume a systematic uncertainty of 10\% for the signal and background samples, and a branching ratio into dark matter of 100\%. 

Assuming 10 (100)~fb$^{-1}$ of luminosity, we are sensitive to couplings $g_{v}<2.5~(1.2)$ in all the  mass range considered, $100~\gev<m_{H(A)}<500~\gev$.
Since the pseudoscalar tend to be more boosted than the scalar hypothesis, we observe stronger bounds for the CP-odd mediator, largely driven by $\slashed{E}_T$. Note that once $m_{H(A)} > 2m_t$, the top-pair decay channel opens, and one would expect the branching ratio to dark matter to drop. This is typically an ${\cal O}(1)$ effect for $g_v \sim g_\chi$, but exact calculation requires an assumption to be made for $g_\chi$.  To maintain our 2D parameter space, in this theory study we continue to assume a 100\% invisible branching ratio.

Also in Figure~\ref{fig:constrain}, we show the 95\% CL exclusion on the coupling $g_v$ as a function of mediator mass $m_{H(A)}$ from the non-collider experimental searches for dark matter: direct  and indirect detection. The scalar mediator $H$ can induce spin-independent scattering in direct detection experiments. Currently, the most constraining spin-independent direct detection limits for dark matter $m_\chi \gtrsim 10$~GeV come from the LUX experiment \cite{Akerib:2013tjd}. Pseudoscalar mediators do not result in direct detection scattering cross sections which are not velocity- or momentum-suppressed. Indirect detection bounds are extracted from the combined dwarf galaxy analysis in the $b\bar{b}$ channel using data from the {\it Fermi} Large Area Telescope (LAT) \cite{Ackermann:2015zua}; such bounds apply to the pseudoscalar mediators $A$, as scalar mediators have thermally averaged annihilation cross sections which are proportional to velocity-squared, which is essentially zero in the Universe today. Finally, in both CP assignments we show the value of $g_v$ required for a thermal relic, assuming $\langle \sigma v \rangle = 3 \times 10^{-26}~$cm$^3$/s in the early Universe. For both models, we set $g_\chi = g_v$ in order to allow for direct comparisons with the collider bounds. Note that varying this assumption can change the relative strength of the collider and non-collider bounds. 

As can be seen, under these assumptions, the LHC bounds with 10~fb$^{-1}$ of data are weaker than the existing LUX constraints. With the larger data set, the LHC can outperform direct detection at large mediator masses. Pseudoscalar mediators are much more difficult to constrain without collider bounds, as even the smaller integrated luminosity we consider is sufficient to exceed the constraints set by indirect detection. Regardless of the CP-structure of the mediator, the requirements for a successful thermal relic seem to require values of $g_v$ much larger than those allowed by existing non-LHC experiments. However, some caution should be applied to this conclusion, as the {\it Simplified Model} framework by construction does not aim to be a complete model of the dark sector physics. For example, additional annihilation modes, not captured in our assumption of a single mediator connecting dark matter to the Standard Model, can allow for smaller values of $g_v$ while still resulting in large enough thermal cross sections.\medskip

\begin{figure}[t!]
\centering
\includegraphics[width=1.\columnwidth]{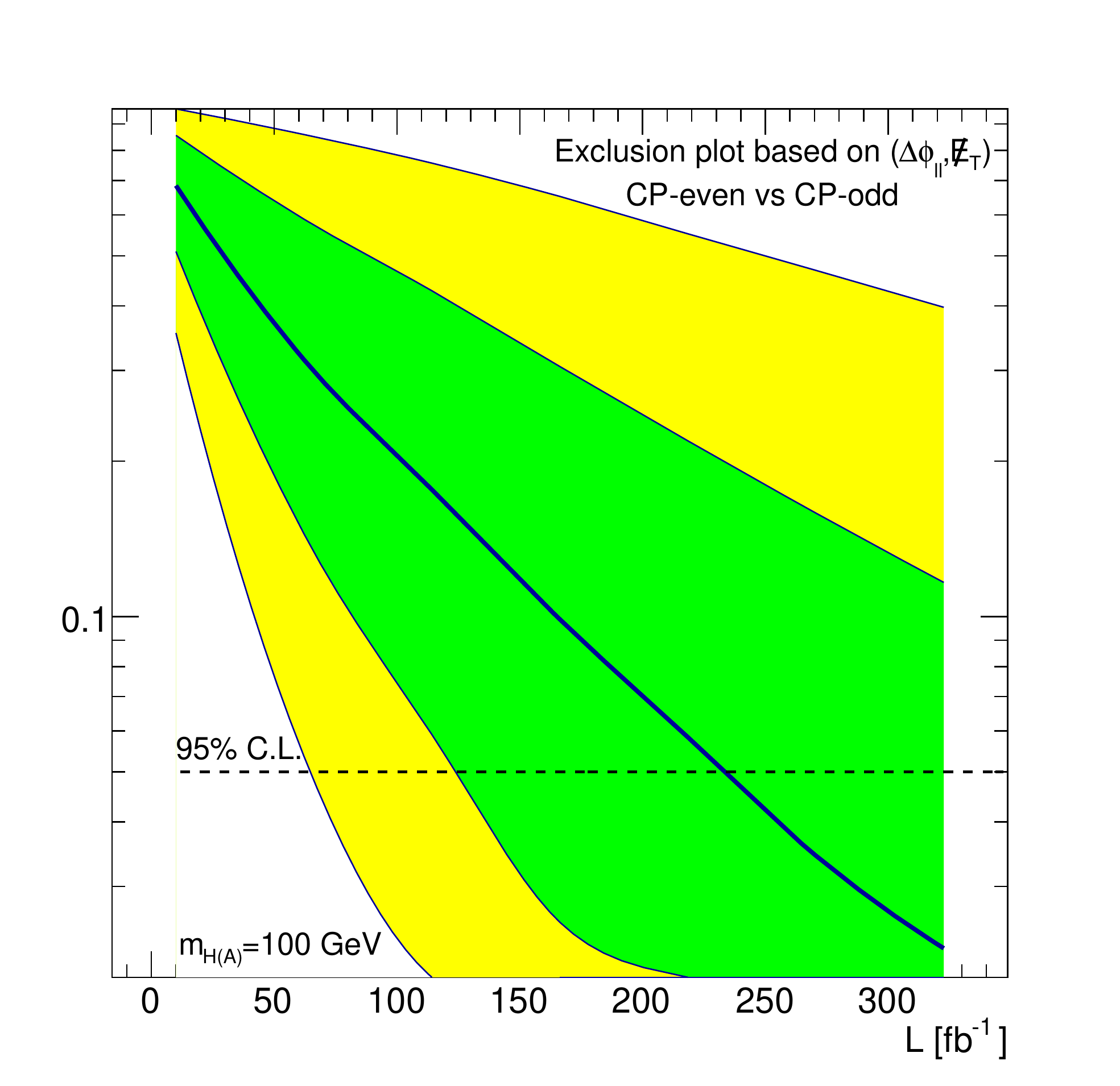}
\caption{Confidence level for disentangling the CP-even from the CP-odd  Dark matter mediators. We assume $g_v=1$ and $m_{H(A)}=100$~GeV with
 $BR(H,A\rightarrow \rm{inv.})=1$. The log-likelihood test is based on the two dimensional distributions $(\Delta\phi_{\ell\ell},\slashed{E}_T)$.
\label{fig:distinguish}}
\end{figure}

As shown, the azimuthal angle $\Delta\phi_{\ell\ell}$ can be used to boost the experimental sensitivity to dark matter produced via a spin-0 mediator. If we further assume discovery of a signal at the LHC consistent with dark matter production, this variable can also be used to distinguish between the scalar and pseudoscalar interactions. In Figure~\ref{fig:distinguish}, we show the estimated CL$_{\rm S}$ value for distinguishing a 100~GeV scalar mediator from a pseudoscalar, as a function of integrated LHC luminosity, using the $\Delta \phi_{\ell\ell}$ and $\slashed{E}_T$ variables. Here again, we assume $g_v = g_\chi = 1$. Under this set of assumptions, the two models can be distinguished at 95\% CL with ${\cal L} \sim 250$~fb$^{-1}$.

As we have shown, the $\Delta\phi_{\ell\ell}$ variable would allow for direct measurement of the CP-structure of the spin-0 mediator between top quarks and dark matter -- to our knowledge the first study to do so. This measurement would be orthogonal to the information that could be inferred from  (non-)detection in direct or indirect detection experiments. We note that the azimuthal angle between jets in vector boson fusion is also known to encode information about the CP-properties of new physics \cite{Eboli:2000ze,Plehn:2001nj,Buckley:2010jv,Englert:2012xt,Klamke:2007cu, Buckley:2014fqa,spano_gf}, and one might look to that channel for a similar measurement, producing a $\Phi$ mediator through a top-loop induced gluon fusion with two forward jets. It should be noted that this channel may suffer from large backgrounds that are difficult to remove, and that the measurement of CP in that case would be a measurement of a loop-induced coupling, rather than the tree-level coupling this work is sensitive to~\cite{Buschmann:2014sia}. We will consider the CP measurement in the vector boson channel in a future work.

In the event of an excess of events seen in the $t\bar{t}+\slashed{E}_T$ channel at the LHC Run-II,  measurements such as the one described here will become a critical part of deciphering the unknown physics of the dark sector. The knowledge gained from CP-measurements at colliders can be integrated with additional signals (or lack thereof) from direct and indirect detection in order to move beyond the {\it Simplified Model} framework into a coherent picture of the particles involved and the nature of their interactions. 



\end{document}